\documentclass[twocolumn,showpacs,preprintnumbers,amsmath,amssymb,nofootinbib]{revtex4-2}
\makeatletter
\auto@bib@empty
\let\auto@bib@innerbib\@empty
\makeatother
\begin{document}

\title{Root systems and Minimal RG flows to $c=1$ CFTs}

\newcommand{\sharedaffil}{\vspace{3pt}\\${}^{\dagger}$Department of Physics, Ewha Womans University, Seoul 03760, Korea\\
${}^{\sharp}$University of Bologna, 
Department of Physics and Astronomy, 40126 Bologna, Italy\\
${}^{\flat}$I.N.F.N., Sezione di Bologna, 40126 Bologna, Italy}

\author{Changrim Ahn\,$^{\dagger}$}
\affiliation{\sharedaffil}
\author{Sara Kostadinovic\,$^{\sharp,\flat}$}
\affiliation{\sharedaffil}
\author{Francesco Ravanini\,$^{\sharp,\flat}$}
\affiliation{\sharedaffil}

\begin{abstract}
\centering\begin{minipage}{\dimexpr\paperwidth-6.0cm}
\
We derive exact massless $S$-matrices from algebraic root-system geometry and derive thermodynamic Bethe ansatz (TBA) equations that generate minimal renormalization group (RG) flows.
We show how P. Dorey's geometric construction, originally developed for massive $S$-matrices, can be extended to massless cases and prove that these new $S$-matrices straightforwardly satisfy the bootstrap equations.
We focus on minimal $D_n$ coset CFTs, which are $c=1$ free-boson CFTs with compactification radius $\sqrt{n/2}$, and find their UV completions by analyzing the TBA.

\end{minipage}
\end{abstract}
\maketitle

\section{Introduction}
Symmetry plays a fundamental role in understanding both conformal field theories (CFTs), which describe quantum field theories (QFTs) at renormalization-group (RG) fixed points, and integrable QFTs, obtained by adding special conformal-symmetry-breaking fields to CFTs.
In this Letter, we show how mathematical symmetry can determine the RG flows described by these QFTs, which interpolate between two RG fixed points, the UV and IR CFTs.

In two dimensions, conformal symmetry provides exceptionally powerful tools for characterizing fixed points through their central charges, operator spectra, and correlation functions. Consequently, researchers have studied RG flows between two-dimensional CFTs using various analytical and numerical methods \cite{ZamRG, Zam_c}.

One recent approach is based on exact thermodynamic Bethe ansatz (TBA) equations \cite{alyosha_tba,alyosha_tba_flow} derived from massless $S$-matrices \cite{AhnLeC}.
If a CFT deformed by a relevant operator maintains integrability, it can be described by factorizable $S$-matrices among massive excitations \cite{BCDS, KlaMel}.
The perturbed CFT remains integrable even with additional irrelevant higher energy-momentum tensors, which modify the $S$-matrices by multiplying them by CDD-like scalar functions \cite{SmiZam}.

In the limit where the coefficient of the relevant deformation vanishes, the integrable QFT is described by massless particles in two chiralities: right-moving ($R$) and left-moving ($L$).
The $S$-matrices $S^{RL}_{ij}$ between $R_i$ and $L_j$ particles (with particle species indices $i,j$) are non-trivial functions of the rapidity $\theta$ satisfying the crossing-unitarity condition, explicitly depending on the additional irrelevant operators, while scattering between particles of the same chirality, $S^{RR}_{ij}$ and $S^{LL}_{ij}$, remains the same as those of the original massive $S$-matrices.

RG flows are generated by thermodynamic Bethe ansatz (TBA) systems derived from these $S$-matrices.
Even without detailed $S$-matrices, one can restrict UV-complete theories by imposing finite central charges on the plateau equations, which are the UV limit of the TBA.
Solving these algebraic equations identifies UV-complete theories and three types of RG flows: minimal, diagonal, and saturated \cite{AhnLeC}.
\footnote{As discovered recently for the $E_8$ algebra \cite{AhnKim}, some isolated exceptional RG flows may be added to the list, which we will not discuss here.}

Among these, the minimal flows interpolate from $[G]_2$ to $[G]_1$, where $[G]_k\equiv G_1\times G_k/G_{k+1}$ are coset CFTs and $G_k$ denotes the WZW CFT at level $k$, with $G=A,\,D,\,E$.
Although these flows were already noticed based on guessed TBA systems in \cite{Ravanini},
they have remained mere conjectures since they have not yet been derived from exact $S$-matrices.
The $S$-matrices that generate the minimal RG flows are most important because
$S$-matrices for other types of RG flows can be obtained straightforwardly from the minimal ones; the diagonal by $S_{ij}^{RL}=1/S_{ij}^{RR}$ and the saturated by squares of the minimal $S$-matrices.

The minimal massless $S$-matrices have been found for $G=A_n,\,E_8$ in \cite{AhnKim, Ahn1} by solving the bootstrap equations.
In this Letter, we focus on the massless $S$-matrices of the integrable deformations of the $[D_n]_1$ coset CFTs.  
This CFT series is interesting because it has central charge $c=1$ with compactification radius $\sqrt{n/2}$ \cite{ginsparg}.
Finding UV CFTs and exact RG flows that end at $c=1$ is important for understanding UV completions of string theories, for example.

Instead of solving the bootstrap equations, we take a different approach for $G=D_n$. 
We apply P. Dorey's geometric construction for the massive $S$-matrices in terms of root vectors and Weyl elements of the $D_n$ Lie algebra \cite{Dorey}. 
We show that this formalism extends to the R--L scattering amplitudes $S_{ij}^{RL}$, which generate the minimal RG flows.
In this way, we can understand all the minimal $S$-matrices that generate the RG flows $[G]_2\to [G]_1$ for all simply-laced algebras $G$ in terms of Lie-algebraic symmetries.

\section{Dorey's Geometric Construction}
P. Dorey has constructed diagonal $S$-matrices in terms of Weyl group orbits of root systems of simply-laced Lie algebras 
\cite{BCDS}. 
In this approach, the bootstrap equations become identities when the root vectors form bound-state triangle relations.
We begin with a brief review of this construction, which highlights the beauty of mathematical symmetries.

Let $w$ be the Coxeter element of a simply-laced algebra with rank $r$, defined by $w=w_r\cdots w_1$ 
in terms of Weyl reflection elements $w_i$'s for each simple root $\alpha_i$.  
A Coxeter orbit $\Gamma_i$ is defined by  
\begin{equation}
 \Gamma_i=\{w^{-p}\phi_i\mid p=0,\ldots,h-1\},
\end{equation}
with dual Coxeter number $h$ and associated with a massive particle $i$.
Here $\phi_i$ is a distinguished positive root defined by
\begin{equation}
 \phi_i=w_{r}w_{r-1}\cdots w_{i+1}\alpha_i.
\end{equation}
The projection of each orbit onto the $s=1$ Coxeter plane is a regular $h$-gon; $w$ acts there as a rotation by $2\pi/h$.  
Thus, relative positions of orbit elements encode physical fusing angles.

Simple roots are split into two subsets corresponding to coloring each node of a Dynkin diagram alternatingly with black nodes $\bullet$ and white nodes $\circ$. 
Dorey's normalization is $u_{ij}\equiv u(\phi_i,\phi_j)$ is given by
\begin{equation}
 u_{ij}=\begin{cases}
 0,& i,j\text{ of the same colour},\\
 -1,& i\in\bullet,\ j\in\circ,\\
 +1,& i\in\circ,\ j\in\bullet.
 \end{cases}
\end{equation}

For roots $\alpha,\beta$, an integer-valued signed angle $u(\alpha,\beta)$ ($\mathrm{mod}\ 2h$ in units of $\pi/h$) between their projections onto the Coxeter plane increases by two units when the Coxeter element $w$ is applied as follows:
\begin{equation}
 u(w\alpha,\beta)=u(\alpha,\beta)+2,
\end{equation}
Using these, all other signed angles between elements of two orbits satisfy
\begin{equation}
 u(w^{-p}\phi_i,w^{-q}\phi_j)=u_{ij}-2p+2q
 =u_{ij}-2(p-q).
\end{equation}

A three-particle coupling is nonzero precisely when there are roots from the corresponding orbits satisfying the root-triangle condition
\begin{equation}
\alpha_{(i)}+\alpha_{(j)}+\alpha_{(k)}=0,
 \qquad \alpha_{(a)}\in\Gamma_a .
 \label{roottriangle}
\end{equation}
The fusion angles are related to the signed angles by
\begin{equation}
U^k_{ij}=\frac{\pi}{h}|u(\alpha_{(i)},\alpha_{(j)})|.
\end{equation}

Dorey's $S$-matrix for the massive particles (and the R-R and L-L amplitudes) can be written as
\begin{equation}
 S_{ij}(\theta)=\prod_{p=0}^{h-1}
 \{2p+1+u_{ij}\}_{+}^{(\lambda_i,w^{-p}\phi_j)},\quad i,j=1,\cdots,n,
\label{doreyS}
\end{equation}
where $\lambda_i$ is a fundamental weight.  
Here, the brackets are defined by
\begin{equation}
\{x\}_{+}\equiv 
 (x+1)_{+} (x-1)_{+},\quad (x)_{+}=\sinh\frac{1}{2}\left(\theta+\frac{i\pi x}{h}\right).
\label{massivebracket}
\end{equation}

The bootstrap equations can be written as
\begin{equation}
S_{k\ell}(\theta)S_{i\ell}(\theta+iU^{j}_{ik})S_{j\ell}(\theta-iU^{i}_{j\ell})=1. 
\label{bsnewform}
\end{equation}
The bootstrap shifts of the three amplitudes are precisely the signed angles between these three projected roots.  
Because a rapidity shift translates the block label, the three factors can be aligned so that, for every Coxeter step $p$, they become the \emph{same} block.  Its total exponent is then
\begin{equation}
 (\lambda_l,w^{-p}\alpha_{(i)})
 +(\lambda_l,w^{-p}\alpha_{(j)})
 +(\lambda_l,w^{-p}\alpha_{(k)})
=0
\label{doreyproof}
\end{equation}
from Eq.\eqref{roottriangle}, which makes the bootstrap equations \eqref{bsnewform} hold.

\section{Minimal $S$-matrices for $[D_{n}]_1$}

The R--L scattering amplitudes $S_{ij}^{RL}$ do satisfy the same bootstrap equations as the R--R and L--L ones. 
Therefore, these $S$-matrices should be represented by a similar root system construction.
The exponent factors in \eqref{doreyS} should be maintained to satisfy the key relation Eq.\eqref{doreyproof} along with the root-triangle condition \eqref{roottriangle}. 
One may consider multiplying an overall positive integer into the exponents. 
However, this is limited to an integer factor of $2$, which produces saturated $S$-matrices; beyond that, UV finiteness is violated.
Therefore, we should keep the exponents the same as in \eqref{doreyS}.

Another key point is that the same curly bracket blocks with a given $p$ in \eqref{doreyS} should appear in \eqref{bsnewform}. 
We should also require that $S_{ij}^{RL}$ has no poles in the physical strip, while $S_{ij}^{RR}$ does.
All these points lead to the minimal $S$-matrices, which are still given by \eqref{doreyS}, but with a new definition of the curly bracket function:
\begin{equation}
\{x\}_{+}\equiv 
 (-x)_{+}=\sinh\frac{1}{2}\left(\theta-\frac{i\pi x}{h}\right).
\label{minimalbracket}
\end{equation}
We have checked explicitly that these root system constructions for the minimal $S$-matrices match 
with those of $A_n$ series \cite{Ahn1} and the $E_8$ one \cite{AhnKim}.

Now we consider the coset CFTs $[D_{n}]_1$, which describe $c=1$ free-boson CFT with 
the compactification radius $\sqrt{n/2}$.
Applying the above $S$-matrices to the deformed $[D_{n}]_1$ coset CFTs, 
we can generate an RG flow from certain UV CFTs to these $c=1$ CFTs with this radius 
in terms of the minimal $S$-matrices.

We can obtain concrete expressions for these $S$-matrices by working explicitly with root vectors in each orbit.
For $n\ge 4$, the minimal $S^{RL}_{ij}$ elements are given by
\begin{eqnarray}
S^{RL}_{ab}&=&\prod_{k=1}^{{\rm min}(a,b)}[a+b+1-2k],
\label{minimalSi}\\
S^{RL}_{an}&=&S^{RL}_{a,n-1}=\prod_{k=1}^{a}(n+a-2k),\\
S^{RL}_{n,n}&=&S^{RL}_{n-1,n-1}=\prod_{k=1}^{\lfloor n/2\rfloor}(4k-3),
\label{minimalSnn}\\
S^{RL}_{n,n-1}&=&S^{RL}_{n-1,n}=\prod_{k=1}^{\lfloor (n-1)/2\rfloor}(4k-1),
\label{minimalSii}
\end{eqnarray}
with $a,b=1,\cdots,n-2$ 
where we use the floor function $\lfloor x\rfloor$, and 
define two conventional bracket functions in terms of the curly ones defined in \eqref{minimalbracket}
\begin{equation}
(x)\equiv\frac{(-x)_+}{(x)_+},\quad
[x]\equiv (x)(h-x),\quad h=2(n-1).
\end{equation}
It is remarkable that the minimal R--L scattering amplitudes \eqref{minimalSnn} and \eqref{minimalSii} between two spinor particles (the $(n-1)$-th and the $n$-th) are different, while those of the massive ones (hence, R--R and L--L) are the same up to a sign.  

\section{Derivation of TBA systems for $[D_{n}]_1$ }

From the explicit expressions for the minimal $S^{RL}_{ij}$ in Eqs.\eqref{minimalSi}-\eqref{minimalSii},
we can write them 
\begin{equation}
S^{RL}_{ij}=\prod_{p=1}^{h-1}(p)^{N_{ij}(p)}, \quad i,j=1,\cdots,n,
\end{equation}
from which we can define $n\times n$ exponent matrices $N(p)$ with $p=1,\cdots,h-1$.
One can easily read off from Eqs.\eqref{minimalSi}-\eqref{minimalSii} that $N(1)=\mathbb{I}$, the identity matrix, and $N(2)=G$, the adjacency matrix.
Furthermore, these matrices satisfy recursion relations
\begin{equation}
N(p+1)+N(p-1)=G\cdot N(p), \quad p=1,\cdots,h-1,
\label{recur}
\end{equation}
with $N(0)=N(h)=0$.
Therefore, all the exponent matrices are polynomials of the adjacency matrix $G$.

The standard procedure of deriving the TBA equations from $S$-matrices leads to so-called raw TBAs where the pseudo-energies for each $n$ species are all coupled. 
To convert these to the universal TBA systems in which the underlying symmetry manifests, the kernels from $S^{RL}_{ij}$ should satisfy the inversion relations 
\cite{ZamADE}.
The Fourier-transformed logarithmic derivatives of the $S^{RL}_{ij}$, in terms of those of the bracket functions $(p)$, are given by
\begin{equation}
\widetilde{\Psi}_{ij}=\sum_{p=1}^{h-1} N_{ij}(p)\left( \frac{q^{h-p}-q^{-h+p}}{q^h-q^{-h}}\right),\quad q=e^{\omega/h}.
\end{equation}
Treating these as matrix elements of $\widetilde{\Psi}$, we can derive an inversion relation
\begin{equation}
\left[(q+q^{-1})\mathbb{I}-G\right]\cdot\widetilde{\Psi}=N(1)=\mathbb{I},
\label{inversion}
\end{equation}
using the recursion relation \eqref{recur}.
We can rewrite this as
\begin{equation}
\left[\mathbb{I}-\widetilde{\varphi}_u\, G\right]\cdot\widetilde{\Psi}=\widetilde{\varphi}_u\mathbb{I},\quad
\widetilde{\varphi}_u=\frac{1}{q+q^{-1}},
\label{inversion2}
\end{equation}
where $\widetilde{\varphi}_u$ is the Fourier transform of the universal kernel.
The $Y$-system can be immediately written as
\begin{equation}
Y^R_i(\theta+\frac{i\pi}{h}) Y^R_i(\theta-\frac{i\pi}{h}) =\frac{\prod_{j=1}^n (1+Y^R_j(\theta))^{G_{ij}}}{(1+1/Y^L_i(\theta))}, 
\label{ysystem}
\end{equation}
with $i=1,\cdots,n$, and similar equations by switching $R\leftrightarrow L$.
Here we have defined $Y_i^{R/L}=\exp\epsilon_i^{R/L}$ in terms of the pseudo-energies of $R/L$-nodes, respectively.
These functional equations can be converted to the conjectured TBA equations \cite{Ravanini} in terms of the universal kernel $\varphi_u$, which generates the RG flows $[D_n]_2\to [D_n]_1$
along the direction of the primary field $\phi_{1,1,{\rm adj}}$ with the conformal dimension $3/(2n+1)$ for each $n\ge 4$.

\section{Summary and Discussion}
In this Letter, we found the minimal $S$-matrices between massless particles in two chiralities for $[D_n]_1$ coset CFTs deformed by a set of higher-energy-momentum irrelevant operators.
These CFTs correspond to $c=1$ free-boson CFTs with compactification radius $\sqrt{n/2}$.
UV-complete CFTs can be connected to each of these $c=1$ CFTs by
the minimal RG flows $[D_n]_2\to [D_n]_1$ generated by the minimal $S$-matrices. 

In addition to the minimal RG flows, other allowed RG flows are also limited to the diagonal RG flows generated by the diagonal $S$-matrices with $S^{RL}_{ij}=1/S^{RR}_{ij}$, whose UV limit leads to generalized $D_n$ parafermions \cite{Gepner}, and the saturated flows generated by 
the square $(S^{RL}_{ij})^2$ of the minimal ones. 
Although we have not addressed the UV completeness issue much here, we confirmed that the minimal, saturated, and diagonal solutions are the only possible solutions to the bootstrap equations for the $D_n$ flows.
We also proved the conjectured TBA systems \cite{Ravanini} using inversion relations derived from the minimal $S$-matrices.

Our main result is to find the minimal $S$-matrices by generalizing P. Dorey's geometric construction of the $S$-matrices in terms of root systems and Weyl elements of the underlying Lie-algebraic symmetries.

It will be interesting to classify all possible massless $S$-matrices in terms of the geometric construction for other integrable QFTs that exhibit RG flows.

\section*{Acknowledgements}

This work was supported by the National Research Foundation of Korea (NRF) grant RS-2026-25472596 (CA) and by the INFN Grant GAST (FR and SK). FR  also acknowledges financial support from the European Union - Next Generation EU through MUR projects PE0000023-NQSTI (Italy).



\begin{thebibliography}{99}
\bibitem{ZamRG}  A. B. Zamolodchikov, 
{\em Renormalization group and perturbation theory about fixed points in two-dimensional field theory}, Yad. Fiz. {\bf 46} (1987) 1819.

\bibitem{Zam_c}  A. B. Zamolodchikov, 
{\em Irreversibility of the flux of the renormalization group in a 2D field theory},
Pis'ma Eksp. Teor. Fiz. {\bf 43} (1986) 565.

\bibitem{alyosha_tba} Al. B. Zamolodchikov,  {\em Thermodynamic Bethe ansatz in Relativistic Models: Scaling 3-state Potts and Lee-Yang Models}, Nucl. Phys. {\bf B342} (1990) 695.

\bibitem{alyosha_tba_flow}Al. B. Zamolodchikov, {\em From Tricritical Ising to Critical Ising By Thermodynamic Bethe ansatz}, Nucl. Phys. {\bf B358} (1991) 524.

\bibitem{AhnLeC} C. Ahn and A. LeClair, {\em On the classification of UV completions of integrable $ T{\overline T}$  deformations of CFT}, JHEP {\bf 2022} (2022) 179, arXiv:2205.10905 [hep-th].

\bibitem{BCDS} H. W. Braden, E. Corrigan, P. E. Dorey, and R. Sasaki, 
{\em Affine Toda field theory and exact S-matrices},  Nucl. Phys. {\bf B338} (1990) 689.

\bibitem{KlaMel}  T. Klassen and E. Melzer,
{\it The thermodynamics of purely elastic scattering theories and conformal perturbation theory},
Nucl. Phys. {\bf B350}  (1991) 635.

\bibitem{SmiZam} F. A. Smirnov and A. B. Zamolodchikov,  {\em On the space of integrable quantum field theories}, Nucl. Phys. {\bf  B915}  (2017) 363, arXiv:1608.05499 [hep-th].

\bibitem{AhnKim} C. Ahn and M. Kim, {\em UV completion of 2D Ising CFT:
a golden $E_8$ massless S-matrix}, arXiv:2606.28783 [hep-th].

\bibitem{Ravanini} F. Ravanini, 
{\it Thermodynamic Bethe Ansatz for $G_k \otimes G_\ell / G_{k + \ell}$ coset models perturbed by their $\phi_{1,1,{\rm Adj}} $ operator}, 
Phys. Lett. {\bf B282} (1992) 73, arXiv:hep-th/9202020.

\bibitem{Ahn1} C. Ahn, {\em Massless $S$-matrix Bootstraps and RG Flows}, JHEP {\bf 2512} (2025) 103, arXiv:2509.20740 [hep-th].

\bibitem{ginsparg} P. Ginsparg, {\em Curiosities at $c=1$}, Nucl. Phys. {\bf B295} (1988) 153.

\bibitem{Dorey} P.~Dorey, \emph{Root Systems and Purely Elastic S-Matrices}, Nucl. Phys. {\bf B358} (1991) 654; \emph{Root Systems and Purely Elastic S-Matrices II}, Nucl. Phys. {\bf B374} (1992) 741.

\bibitem{ZamADE} Al. B. Zamolodchikov, {\it On the thermodynamic Bethe ansatz equations for reflectionless 
ADE scattering theories}, Phys. Lett. {\bf B253} (1991) 391.

\bibitem{Gepner} D. Gepner, New conformal field theories associated with  Lie algebras and their partition functions, 
Nucl. Phys. {\bf B290} [FS20] (1987) 10.


\end{thebibliography}
\end{document}